\documentclass[doublecol]{epl2} 
\usepackage{graphicx}
\usepackage{epstopdf}

\newcommand{\kin}{K_{\textrm{\scriptsize in}}}
\newcommand{\kout}{K_{\textrm{\scriptsize out}}}

\title{Distinguishing humans from computers in the game of go: a complex network approach}
\shorttitle{Distinguishing human and computers in the game of go} %Insert here a short version of the title if it exceeds 70 characters

\author{C. Coquid\'e\inst{1} \and B. Georgeot\inst{1} \and O. Giraud\inst{2}}
\shortauthor{C. Coquid\'e \etal}

\institute{                    
  \inst{1}Laboratoire de Physique Th\'eorique, IRSAMC, Universit\'e de Toulouse, CNRS, UPS, 31062 Toulouse, France\\
\inst{2} LPTMS, CNRS, Univ.~Paris-Sud, Universit\'e Paris-Saclay, 91405 Orsay, France
}


\abstract{We compare complex networks built from the game of go and obtained from databases of human-played games with those obtained from computer-played games. Our investigations show that statistical features of the human-based networks and the computer-based networks differ, and that these differences can be statistically significant on a relatively small number of games using specific estimators. We show that the deterministic or stochastic nature of the computer algorithm playing the game can also be distinguished from these quantities. This can be seen as tool to implement a Turing-like test for go simulators.}

\begin{document}

\maketitle

%\section{Section title}
%Insert here the text.
%See fig.~\ref{fig.1}, table~\ref{tab.1} and eq.~(\ref{eq.1}).
%See also~\cite{b.a,b.b}.
%\begin{equation}
%\label{eq.1}
%0\neq1
%\end{equation}

%%%%%%%%%%%%%%%%%%%%%%%%%%%%%%%%%%%%%%%%%%%%%%%%%%%%%%%%%%%%%%%%%%%%%%%%%%%%%%%%%%%
\section{Introduction}
%%%%%%%%%%%%%%%%%%%%%%%%%%%%%%%%%%%%%%%%%%%%%%%%%%%%%%%%%%%%%%%%%%%%%%%%%%%%%%%%%%%

Computers are more and more present in everyday life, and they often perform tasks that were previously reserved to human beings. In particular, the raise of Artificial Intelligence in recent years showed that many situations of decision-making can be handled by computers in a way comparable to or more efficient than that of humans. However, the processes used by computers are often very different from the ones used by human beings. These different processes can affect the decision-making in ways that are difficult to assess but should be explored to better understand the limitations and advantages of the computer approach. A particularly spectacular way of testing these differences was put forward by Alan Turing: in order to distinguish a human from a computer one could ask a person to dialog with both anonymously and try to assess which one is the biological agent. As the question of human-computer interaction gets more pregnant, there is an ever growing need to understand these differences \cite{hci14}.  Many complex problems can illustrate the deep differences between human reasoning and the computer approach.  Board games such as chess or go, which are perfect-information zero-sum games, provide an interesting testbed for such investigations. The complexity of these games is such that computers cannot use brute force, as in complex decision making, and have to rely on refined algorithms from Artificial Intelligence. Indeed, the number of legal positions is about $10^{50}$ in chess and $10^{171}$ in go \cite{TroFar07}, and the number of possible games of go was recently estimated to be at least $10^{10^{108}}$ \cite{WalTro16}. This makes any exhaustive analysis impossible, even for machines, and pure computer power is not enough to beat humans. Indeed, the most recent program of go simulation AlphaGo \cite{alphago16} used state of the art tools such as deep learning neural networks in order to beat world champions.

Various approaches were considered to overcome the vastness of configuration space. A cornerstone of the computer approach to board games is a statistical physics treatment of game features. A first possibility is to explore the tree of all games stochastically, an approach which allowed for instance to investigate the topological structure of the state space of chess \cite{sequencingchess16}. A second option is to consider only opening sequences in the game tree. This allowed e.g.~to identify Zipf's law in the tree of openings in chess \cite{chess} and in go \cite{Weiqi15}. A third possibility is to restrict oneself to local features of the game, by considering only local patterns. This approach was taken for instance in \cite{LiuDouLu08}, where the frequency distribution of patterns in professional go game records was investigated.

Local patterns play an essential part in the most recent approaches to computer go simulators \cite{stern2006bayesian}. Pioneering software was based on deterministic algorithms \cite{computers,BouCaz01}. Today, computer algorithms implement Monte-Carlo go \cite{MonteCarloGo, Mogo} or Monte-Carlo tree search techniques \cite{progressive, Brown12, Cou07, GelKoc12}, which are based on a statistical approach : typically, the value of each move is estimated by playing a game at random until its end and by assigning to the move the average winning probability. The random part relies on a playout policy which tells how to weight each probabilistic move. Such a playout policy rests on properties of local features, e.g.~3$\times 3$ patterns with atari status \cite{Coulom2}. The most recent computer go approaches such as in AlphaGo \cite{alphago16}, which famously defeated a world champion in 2016 and 2017, also incorporates local pattern-based features such as $3\times 3$ patterns and diamond-shaped patterns.

In the present work, we investigate the differences between human and computer players of go using statistical properties of complex networks built from local patterns of the game. We will consider networks whose nodes correspond to patterns describing the local situation on the $19\times 19$ goban (board). In the original setting \cite{goGG}, we introduced a network based on $3\times 3$ patterns of moves. We then extended it to larger, diamond-shaped patterns and explored the community structure of the network for human players \cite{goKGG}. Here we will focus on the differences between networks obtained from games played by humans and games played by computers. This study is new to the best of our knowledge. In a parallel way, there have already been previous studies  to distinguish amateur and professional human players by looking at statistical differences between their games. For instance, professional moves in a fixed region of the goban were shown to be less predictable than amateur ones and this predictability turned out to evolve as a function of the degree of expertise of the professional \cite{Harre11}. Differences between amateur players of different levels were also identified in \cite{goKGG}. Here we will show that there are clear differences between complex networks based on human games and those based on computer games. These differences, which appear at a statistical level, can be seen as a signature of the nature of players involved in the game, and reveal the different processes and strategies at work. We will specify which quantities can be used to detect these differences, and how large a sample of games should be for them to be statistically significant.  Additionnally, we will show that this technique allows to distinguish between computer games played with different types of algorithms.

%%%%%%%%%%%%%%%%%%%%%%%%%%%%%%%%%%%%%%%%%%%%%%%%%%%%%%%%%%%%%%%%%%%%%%%%%%%%%%%%%%%
\section{Construction of the networks}
%%%%%%%%%%%%%%%%%%%%%%%%%%%%%%%%%%%%%%%%%%%%%%%%%%%%%%%%%%%%%%%%%%%%%%%%%%%%%%%%%%%

Our network describing local moves in the game of go is constructed in the following way \cite{goGG}. Nodes correspond to $3\times 3$ intersection patterns in the $19\times 19$ goban with an empty intersection at its centre. Since an intersection can be empty, black or white, there are $3^8$ such patterns. Taking into account the existence of borders and corners on the goban, and considering as identical the patterns equivalent under any symmetry of the square as well as colour swap, we end up with $N=1107$ non-equivalent configurations, which are the nodes of our graph. Let $i$ and $j$ be two given nodes. In the course of a game, it may happen that some player plays at a position $(a,b)$ which is the centre of the pattern labeled by $i$, and that some player (possibly the same) plays later in the same game at some position $(a',b')$ which is the centre of the pattern labeled by $j$. If this happens in such a way that the distance between $(a,b)$ and $(a',b')$ is smaller than some fixed distance $d_s$, and that the move at $(a',b')$ is the first one to be played at a distance less than $d_s$ since $(a,b)$ has been played, then we put a directed link between nodes $i$ and $j$. Since part of the go game corresponds to local fights, the distance $d_s$ allows to connect moves that are most likely to be strategically related. Following \cite{goGG} we choose this strategic distance to be $d_{s}=4$. We thus construct from a database a weighted directed network, where the weight is given by the number of occurences of the link in the games of the database. 

In what follows, we will use three databases. The first one corresponds to 8000 games played by amateur humans, and is available online in sgf format \cite{database}. The two other databases correspond to games played by computer programs, using either a deterministic approach or a Monte-Carlo approach. To our knowledge, there is no freely available database for computer games, and therefore we opted for free go simulators. As a deterministic computer player we chose the software Gnugo \cite{gnugo}. Although this program is relatively weak compared to more recent computer programs, it is easy to handle and a seed taken as an input number in the program allows to deterministically reproduce a game. Using 8000 different seeds and letting the program play against itself we constructed a database of 8000 distinct computer-generated games. As a computer player implementing the Monte-Carlo approach we chose the software Fuego \cite{fuego}, placing very well in computer go tournaments in the past few years, with which we generated a database of 8000 games. These databases allowed us to construct three distinct networks, one from the human database and one for each computer-generated one. In order to investigate the role of the database size, we also consider graphs constructed from smaller subsets of these databases (with networks constructed from 1000 to 8000 games).

%%%%%%%%%%%%%%%%%%%%%%%%%%%%%%%%%%%%%%%%%%%%%%%%%%%%%%%%%%%%%%%%%%%%%%%%%%%%%%%%%%
\section{General structure of the networks}
%%%%%%%%%%%%%%%%%%%%%%%%%%%%%%%%%%%%%%%%%%%%%%%%%%%%%%%%%%%%%%%%%%%%%%%%%%%%%%%%%%%

\begin{figure}
\begin{center}
\includegraphics*[width=0.99\linewidth]{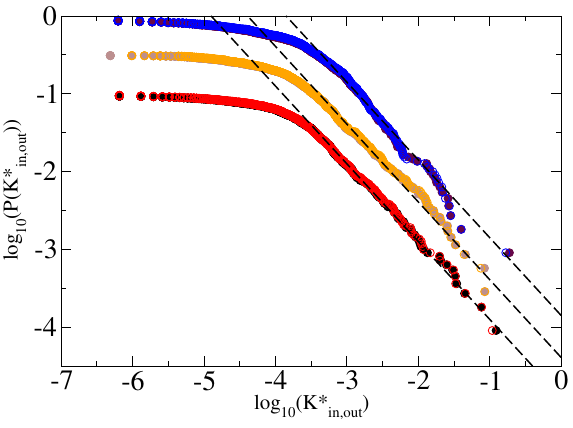}
\end{center}
\caption{Integrated distribution of ingoing links $\kin^{*}=\kin/k_{\textrm{tot}}$ and outgoing links $\kout^{*}=\kout/k_{\textrm{tot}}$. Here $k_{\textrm{tot}}$ is the total number of links in the network (from top to bottom $1 589 729$, $2 046 260$ and $1 527 421$). $P(K^{*})$ is defined as the proportion of nodes having more than $K=k_{\textrm{tot}}K^{*}$ links. From top to bottom, deterministic computer (Gnugo, empty blue $\kin^{*}$, filled maroon $\kout^{*}$), Monte-Carlo computer (Fuego, empty orange $\kin^{*}$, filled grey $\kout^{*}$) and humans (empty red $\kin^{*}$, filled black $\kout^{*}$), shifted down by respectively $0$, $-1/2$ and $-1$ for clarity. The leftmost point corresponds to the nodes with minimal number of links $k_{\textrm{min}}$ (here $k_{\textrm{min}}=1$ or $2$), with abscissa $k_{\textrm{min}}/k_{\textrm{tot}}$ and ordinate $1-N_{0}/N$, $N_{0}$ being the number of nodes with no link. The righmost point corresponds to the node with maximal number of links $k_{\textrm{max}}$ (which happens to be also the node with highest PageRank shown in Fig.~\ref{figPR}), with abscissa $k_{\textrm{max}}/k_{\textrm{tot}}$ and ordinate $1/N$. The networks are all built from 8000 games. Black dashed lines have slope $-1$. \label{figlinks}}
\end{figure}

We first investigate the general structure of the three networks built from all 8000 games for each database. %The network built from human-played games is similar to the one discussed in \cite{goGG,goKGG}. 
Taking into account the degeneracies of the links, each node has a total of $\kin$ incoming links and $\kout$ outgoing links. The (normalized) integrated distribution of $\kin$ and $\kout$, displayed in Fig.~\ref{figlinks} for each network, shows that general features are similar. In all cases, the distribution of outgoing links is very similar to the distribution of ingoing links. This symmetry is due to the fact that the way of constructing the networks from sequentially played games ensures that in most cases an ingoing link is followed by an outgoing link to the next move. The distributions of links are close to power-law distributions, with a decrease in $1/K^{\gamma}$ with $\gamma \approx 1.0$. Networks displaying such a power-law scaling of the degree distribution have been called scale-free networks \cite{AB99}. Many real-world networks (from ecological webs to social networks) possess this property, with an exponent  $\gamma$ typically around 1 (see Table II of \cite{AB02}). Our networks belong to this class, which indicates the presence of hubs (patterns with large numbers of incoming or outgoing links), and more generally a hierarchical structure between patterns appearing very commonly and others which are scarce in the database.

While the distributions for the three networks are very similar, the power-law scaling ends (on the right of the plot) at a smaller value of $K$ in the case of both computer go networks, with strong oscillations. The rightmost points correspond to hubs. The figure shows that such hubs are slightly rarer in the computer case than in the human one. This may indicate that certain moves are preferred by some human players independently of the global strategy, while computers play in a more even fashion.   However, at the level of these distributions the differences between databases seem too weak to lead to reliable indicators.

%%%%%%%%%%%%%%%%%%%%%%%%%%%%%%%%%%%%%%%%%%%%%%%%%%%%%%%%%%%%%%%%%%%%%%%%%%%%%%%%%%%
\section{PageRank}
%%%%%%%%%%%%%%%%%%%%%%%%%%%%%%%%%%%%%%%%%%%%%%%%%%%%%%%%%%%%%%%%%%%%%%%%%%%%%%%%%%%

Each directed network constructed above can be described by its $N\times N$ weighted adjacency matrix $(A_{ij})_{1\leq i,j\leq N}$, with $N=1107$, such that $A_{ij}$ is the number of directed links between $i$ and $j$ as encountered in the database. The PageRank vector, that will be defined below, allows to take into account the network structure in order to rank all nodes according to their significance within the network. It allows to go beyond the mere frequency ranking of the nodes, where nodes would be ordered by the frequency of their occurence in the database. Physically speaking, the significance of a node is determined by the average time that would be spent on it by a random walker moving on the network by one step per time unit and choosing a neighouring node at random with a probability proportional to the number of links to this neighbour. Such a walker would play a virtual game where, at each step, it can play any move authorized by the network, with some probability given by the network. The PageRank vector assigns to any node $i$ a nonnegative value corresponding to the equilibrium probability of finding this virtual player on node $i$. 

\begin{figure}
\begin{center}
\includegraphics*[width=.99\linewidth]{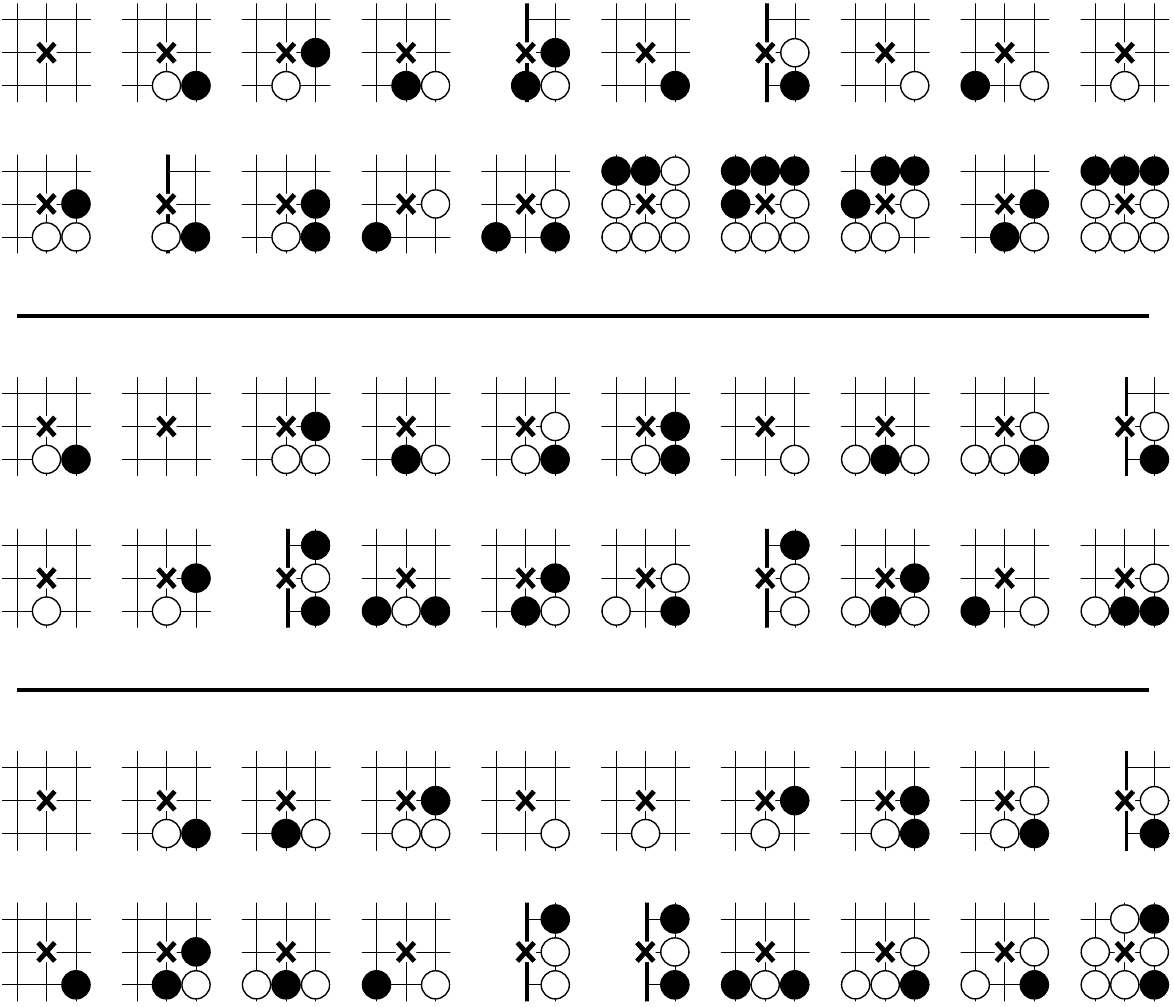}
\end{center}
\caption{First 20 patterns as ranked by the PageRank vector, for the
  networks generated by Gnugo (top), Fuego (middle) and humans (bottom), each built from 8000 games, for $\alpha=0.85$ (see text). Black plays at the cross. \label{figPR}}
\end{figure}

More precisely, the PageRank vector is obtained from the Google matrix $G$, defined as $G_{ij}=\alpha S_{ij}+(1-\alpha)/N$, with $S$ the matrix obtained by normalizing the weighted adjacency matrix so that each column sums up to 1 (any column of 0 being replaced by a column of $1/N$), and $\alpha$ is some parameter in $[0,1]$. Since $G$ is a stochastic matrix (all its columns sum up to 1), there is a vector $p$ such that $Gp=p$ and $p_i\geq 0$ for $1\leq i\leq N$. This right eigenvector of $G$, associated with the eigenvalue 1, is called the PageRank vector. We can then define the corresponding ranking vector $(a_k)_{1\leq k\leq N}$, with $1\leq a_k\leq N$, as the permutation of integers from 1 to $N$ obtained by ranking nodes in decreasing order according to the entries $p_i$ of the PageRank vector, namely $p_{a_1}\geq p_{a_2}\geq \ldots\geq p_{a_N}$. As an illustration, we show in Fig.~\ref{figPR}, for each network, the 20 nodes $a_1,\ldots,a_{20}$ with largest PageRank vector entry $p_i$. The distinction clearly appears. For instance, among the 20 entries of the human network only 12 appear in the Gnugo Pagerank vector (18 in the Fuego one).

\begin{figure}
\begin{center}
\includegraphics*[width=.99\linewidth]{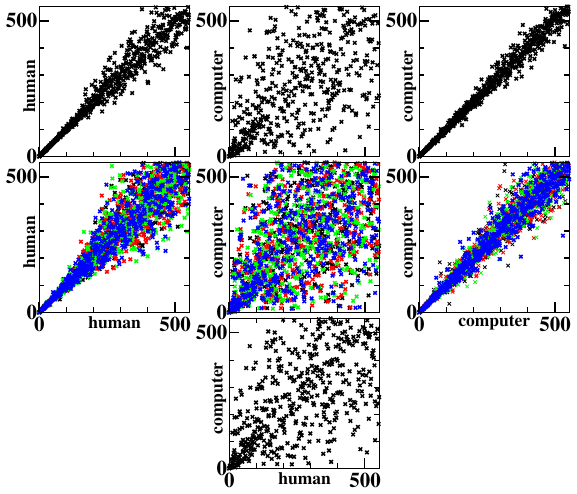}
\end{center}
%\vskip 0.5cm
%\hskip 1.8cm
%\includegraphics*[width=.5\linewidth]{correlationpaperbonuswith8000games-epl.eps}
\caption{PageRank-PageRank correlation for networks built from subsets of
  the database of computer (Gnugo) or human games ($\alpha=0.85$). First line: databases are split into 2 groups of 4000 games, yielding two distinct networks, and correlator between the two PageRanks is plotted. Second line: database is split into 8 groups of 1000, and 4 correlators are plotted: group 1 vs 2 (black) 3 vs 4 (red), 5 vs 6 (green), 7 vs 8 (blue). Middle column corresponds to human vs computer, left column to human vs human, right column to computer vs computer. Bottom panel: Same for networks constructed from the whole database (8000 human vs 8000 computer games).  Nodes are ranked according to the PageRank of each network. Only the $\lfloor N/2\rfloor=553$ first nodes are plotted. \label{correlPR}} 
\end{figure}

In order to quantify more accurately the discrepancy between the human and
computer PageRank vectors, and between PageRank vectors obtained for different database sizes, we consider the correlations between their associated ranking vectors. If $p$ and $q$ are two PageRank vectors, let $A=(a_k)_{1\leq k\leq N}$ and $B=(b_k)_{1\leq k\leq N}$ be their respective ranking vectors, with $1\leq a_k,b_k\leq N$. The correlations are estimated from the discrepancy between pairs $(a_k,b_k)$ and the line $y=x$. As an illustration, such correlation plots are shown in Fig.~\ref{correlPR}, where pairs $(a_k,b_k)$  are plotted for the Gnugo and human databases. While correlation between two human PageRanks or two Gnugo PageRanks is quite good, the correlation between human and computer-generated networks is very poor. This observation does not depend on the choice or the size of the database: indeed, as appears in Fig.~\ref{correlPR}, several different databases of different sizes all give comparable results. In order to be more quantitative, we introduce the dispersion
\begin{equation}
\label{sigmadef}
\sigma(A,B)=\left(\frac{\sum_{k=1}^{\lfloor N/2 \rfloor} (a_k-b_k)^{2}}{\lfloor N/2 \rfloor} \right)^{1/2},
\end{equation}
where we restrict ourselves to the first half of the entries, corresponding to the largest
values of the $p_i$ (this truncation to $N/2$ amounts to neglecting entries smaller than $p_{a_{554}}$, which, for all samples and database sizes investigated, is of order $\simeq 3.10^{-4}$ for a PageRank vector normalized by $\sum_ip_i=1$). The dispersion gives the (quadratic) mean distance from perfect correlation function $y=x$  to
points $(a_k,b_k)$ in the plot of Fig.~\ref{correlPR}, with  for random permutations $\sigma \approx 450 $ on average. In the case of two groups of 4000 games, the human-human dispersion is 43.66, the computer-computer one is 24.04, while the human-computer one is 192.58. A similar discrepancy can be measured for the 1000 game groups, with a dispersion $\sigma$ (averaged over the different samples) given by {$\sigma = 66.71 $} for human-human , {$\sigma = 44.14 $} for computer-computer, and {$\sigma = 199.27 $} for human-computer. The plot at the bottom of Fig.~\ref{correlPR} is a PageRank correlation plot between human and computer with the whole database of 8000 games for each, giving $\sigma = 193.48$. 

%%%%%%%%%%%%%%%%%%%%%%%%%%%%%%%%%%%%%%%%%%%%%%%%%%%%%%%%%%%%%%%%%%%%%%%%%%%%%%%%%%%
\section{Spectrum of the Google matrix}
%%%%%%%%%%%%%%%%%%%%%%%%%%%%%%%%%%%%%%%%%%%%%%%%%%%%%%%%%%%%%%%%%%%%%%%%%%%%%%%%%%%

\begin{figure}
\begin{center}
\includegraphics*[width=0.99\linewidth]{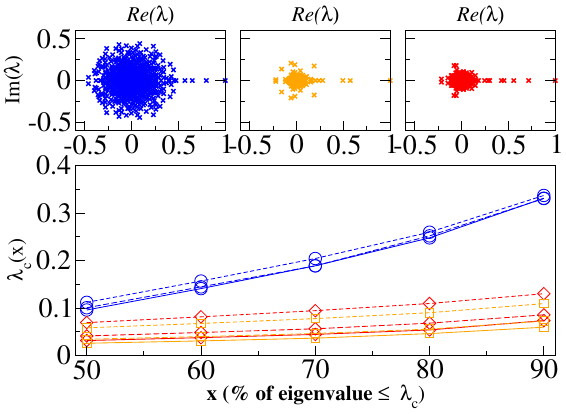}
\end{center}
\caption{Top row : spectrum of the Google matrix for $\alpha=1$ (see text) for the network generated by computers (Gnugo left, Fuego middle) and humans (right). Main plot : $\lambda_{c}(x)$ for $x = 50, 60, 70, 80, 90$ (see text), for Gnugo (blue circles), Fuego (orange squares) and humans (red diamonds), averaged over $m$ networks built from 1000 games ($m=8$, dashed line), 4000 games ($m=2$, long dashed line) and 8000 games ($m=1$, full line).\label{figlambda}} 
\end{figure}

The PageRank vector is the right eigenvector of $G$ associated with the largest eigenvalue $\lambda=1$. It already shows some clear differences between the networks built by computer-played games and human-played games. We now turn to subsequent eigenvalues and eigenvectors. In Fig.~\ref{figlambda} we display the distribution of eigenvalues of the Google matrix in the complex plane for the three networks of 8000 games. Properties of the matrix impose that all eigenvalues lie inside the unit disk, with one of them (associated with the PageRank) exactly at 1, and that complex eigenvalues occur in conjugated pairs. The spectra obtained from different networks are clearly very different: eigenvalues for the network built from computer-played games using Gnugo are much less concentrated around zero, with many eigenvalues at a distance 0.2-0.6 from zero which are absent in the other networks. Moreover, while the bulk of eigenvalues looks similar for games played by Fuego or by humans, many outlying eigenvalues are present in the case of Fuego.
To make these observations more quantitative, we plot in the main panel of Fig.~\ref{figlambda} the radius $\lambda_{c}(x)$ of the minimal circle centred at 0 and containing a certain percentage $x$ of eigenvalues.  The difference between the two behaviours is striking. Considering plots obtained from subsets of the databases, we see that the result is robust: although $\lambda_{c}(x)$ depends much more on the size of the subset used to build the network for Gnugo than for the other two networks, the difference between the plots remains clear.

In fact, the presence of eigenvalues with large absolute value has been related to the existence of parts of the network which are weakly linked to the rest ('communities'). Eigenvalues lying out of the bulk in the Fuego network would mean that there are more communities present in that network than in the human case, and even more in the Gnugo network. Note however that the average number of moves per game is larger for Fuego, which reflects in the total number of links in the network (see caption of Fig.~\ref{figlinks}), and may give a small bias in the comparison. The outlying eigenvalues seem to indicate that the deterministic program, and less markedly the Monte-Carlo one, can create different groups of moves linked to each other and not much linked to the other moves, i.e.~different strategies relatively independent from each other. This can be related with the results displayed in Fig.~\ref{figlinks}, which show the presence of more hubs with large number of links in the network generated by human-played games.

%%%%%%%%%%%%%%%%%%%%%%%%%%%%%%%%%%%%%%%%%%%%%%%%%%%%%%%%%%%%%%%%%%%%%%%%%%%%%%%%%%%
\section{Other eigenvectors of the Google matrix}
%%%%%%%%%%%%%%%%%%%%%%%%%%%%%%%%%%%%%%%%%%%%%%%%%%%%%%%%%%%%%%%%%%%%%%%%%%%%%%%%%%%

\begin{figure}
\begin{center}
\includegraphics*[width=0.95\linewidth]{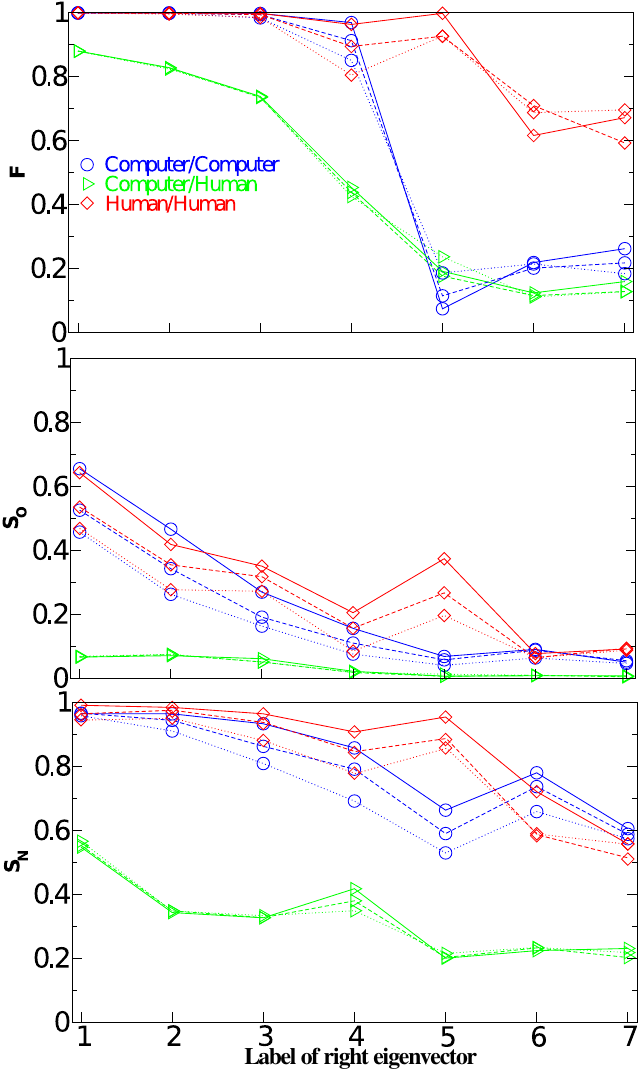}
\end{center}
\caption{Top: Fidelity (\ref{fidelity}) for the 7 first right eigenvectors
  (corresponding to the largest eigenvalues of $G$, including
  the PageRank, for $\alpha=0.85$), for computer (Gnugo) and human networks, in decreasing order of the norm of the associated eigenvalue, for human-human (red diamonds), computer-computer (blue circles), computer-human (green triangles).
  Each point is an average over $N_S$ different choices of pairs of groups.
  Networks are built from groups of 4000 games (solid line, $N_S= 30$), 2000 games
  (dashed line, $N_S=180$) and 1000 games (dotted lines, $N_S=840$) (for computer/human resp. $N_S=120, 480, 1920$). Standard deviation is comparable to symbol size for the 3 first eigenvectors, and is much larger for
  subsequent ones. Middle and bottom: Ordered Vector Similarity (\ref{OVS}) and Non-ordered Vector Similarity (\ref{NVS}) respectively for the 7 first eigenvectors, same conventions and datasets as above. \label{figfidelity}}
\end{figure}

The analysis of the PageRank vector, which is the eigenvector associated with the largest eigenvalue, has shown (see Fig.~\ref{figPR}) that the most significant moves differ between the three networks. When eigenvalues are ordered according to their modulus, as $1=\lambda_1>|\lambda_2|\geq|\lambda_3|\ldots$, the right eigenvectors associated with eigenvalues $\lambda_2,\lambda_3,\ldots$ may be expected to display more refined differences between the networks. 
In order to quantify the difference between eigenvectors of the
Google matrix, we consider two vectors $\phi$ and $\psi$ of
components respectively $\phi_i$ and $\psi_i$, normalized in such a way that
$\sum_i|\phi_i|^2=\sum_i|\psi_i|^2=1$, and we introduce (following the usual definition from quantum mechanics) the fidelity
\begin{equation}
\label{fidelity}
F=|\sum_{i=1}^{N}\phi_i^*\psi_i|,
\end{equation}
where $*$ denotes complex conjugation. The fidelity is $F=1$ for two identical vectors, and $F=0$ for orthogonal ones. In Fig.~\ref{figfidelity} (top panel) we plot the fidelity of the 7 right eigenvectors of $G$ corresponding to the largest eigenvalues $1=\lambda_1>|\lambda_2|\geq|\lambda_3|\geq\ldots\geq |\lambda_7|$. It shows that the fidelity decreases much faster in the computer/human comparison than in both the cases of subgroups of human/human and computer/computer, where it remains very close to 1 for the first 4 eigenvectors, indicating that, remarkably, these eigenvectors only weakly depend on the choice of the data set but strongly on the nature of the players. Interestingly enough, in the computer/computer case there is also a dropoff starting from the fifth eigenvector. This may be due to inversion of close eigenvalues between two realizations of the subgroups.

In order to compare more accurately the eigenvectors at the level of patterns, one can define quantities based on ranking vectors, in line with the ranking of nodes that can be obtained from the PageRank vector. For any eigenvector, we define a ranking vector $A=(a_i)_{1\leq i\leq N}$ with $1\leq a_i\leq N$, where nodes are ordered by decreasing values of the modulus of the components of the vector. We thus define the Ordered Vector Similarity $S_O$, which takes the value 1 if two ranking vectors are identical in their first 30 entries (this choice of cut-off is arbitrary but keeps only the most important nodes). Namely, if $A=(a_i)_{1\leq i\leq N}$ and $B=(b_i)_{1\leq i\leq N}$ are two ranking vectors,  $S_O$ is defined through
\begin{equation}
\label{OVS}
S_O(A,B)=\sum_{i=1}^{30}\frac{f(i)}{30},
\qquad
f(i) = \left\{
    \begin{array}{ll}
        1 & \mbox{if } a_{i}=b_{i}\\
        0 & \mbox{otherwise.}
    \end{array}
\right.
\end{equation}
The similarity $S_O$ gives the proportion of moves which are exactly at the same rank in both ranking vectors within the first 30 entries. This quantity is shown in Fig.~\ref{figfidelity} (middle panel).
Again the data single out the computer/human similarity as being the weakest. However, the choice of the data set affects the results: the dependence on the number of games used to build the networks inside each database is relatively large, and makes the results less statistically significant than for the fidelity. This is due to the fact that some components of the vectors can have very similar values, and a small perturbation can then shuffle the ranking of components. To make this effect less important, we define a Non-ordered Vector Similarity $S_N$ for two vectors A and B through a new similarity function $f_{bis}$:
\begin{equation}
f_{bis}(i) = \left\{
    \begin{array}{ll}
        1 & \mbox{if } \exists\, j \in [1;30] \mbox{ such that } a_{i}=b_{j},  \\
        0 & \mbox{otherwise.}
    \end{array}
\right.
\end{equation}
$S_N$ is thus defined as:
\begin{equation}
\label{NVS}
S_N(A,B)=\sum_{i=1}^{30}\frac{f_{bis}(i)}{30}
\end{equation}
In this quantity, what matters is now the proportion of moves which are common to both lists of the $30$ most important moves, irrespective of their exact rankings through both vectors.
The data are displayed for this quantity on Fig.~\ref{figfidelity} (bottom panel). The dispersion between different choices of subgroup sizes in the same database is now much more reduced, and the results from the human vs computer case are now clearly separated from the ones inside one of the individual databases.

%%%%%%%%%%%%%%%%%%%%%%%%%%%%%%%%%%%%%%%%%%%%%%%%%%%%%%%%%%%%%%%%%%%%%%%%%%%%%%%%%%%
\section{Towards a Turing test for go simulators}
%%%%%%%%%%%%%%%%%%%%%%%%%%%%%%%%%%%%%%%%%%%%%%%%%%%%%%%%%%%%%%%%%%%%%%%%%%%%%%%%%%%

\begin{figure}
\begin{center}
\includegraphics*[width=0.99\linewidth]{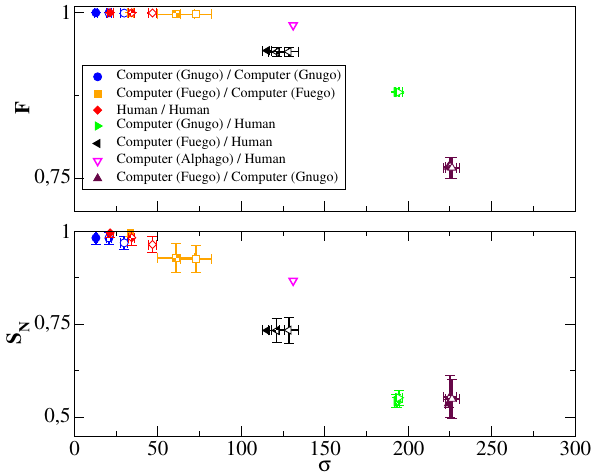}
\end{center}
\caption{Fidelity (\ref{fidelity}) (top) and Non-ordered Vector Similarity (\ref{NVS}) (bottom) of pairs of PageRanks as a function of the dispersion $\sigma$ of (\ref{sigmadef}), for databases of human players and three computer programs: Gnugo, Fuego and AlphaGo. We use $\alpha=0.85$. For each case, one PageRank corresponds to a 8000-game network, and the other one to several choices of networks built from smaller samples; empty symbols correspond to averages over 1000-game networks, checkerboard ones to 2000-game networks, filled ones to 4000-game networks. Averages are made respectively over 240 instances, 120 instances, 60 instances. Error bars are standard deviations of these averages. For AlphaGo, a 50-game network was used. \label{figfidsig}} 
\end{figure}

Figures~\ref{figlinks}--\ref{figfidelity} illustrate the differences between networks built from computer-played games and the ones built from human-played games. These differences are relatively difficult to characterize at the level of the distributions of links of Fig.~\ref{figlinks}. On the other hand, Figs.~\ref{correlPR} and \ref{figfidelity} show that the eigenvectors associated with largest eigenvalues allow more clearly to distinguish between the types of players, with statistical differences
visibly stronger than the ones between the networks built from different subgroups of the same database. This indicates that it may be possible to conceive an indicator to differentiate between a group of human-played games and computer-played games, without any previous knowledge of the players. This could be similar to the famous Turing test of Artificial Intelligence, where a person tries to differentiate a human from a computer from answers to questions, without prior knowledge of which interlocutor is human. In our case, confronted with databases of games from both types of players, it could be possible to differentiate the human from the computer from statistical tests on the network. To construct such an indicator, we focus on the PageRank which corresponds to the largest eigenvalue of the Google matrix, and use the three quantities which distinguish best the different types of players, namely
the fidelity $F$, Non-ordered Vector Similarity $S_N$ and dispersion $\sigma$. The first two quantities describe discrepancies mostly for the largest values of the PageRank, while the dispersion is dominated by intermediate values (see Fig.~\ref{correlPR}). In order to synthetize the results from these two kinds of quantities, we present in Fig.~\ref{figfidsig} the pairs $(F,\sigma)$ and the pairs $(S_N,\sigma)$ for PageRank vectors constructed from games played by humans or computers.

The data displayed in Fig.~\ref{figfidsig} show that there is some variability of these quantities if subgroups from the same databases are compared, indicated by the error bars. However, the difference between the computer- and
human-generated networks is much larger than this variability, indicating that there is a statistically significant difference between them. Interestingly enough, it is also possible to distinguish between the different types of algorithms used in the computer games: 
differences between Fuego and Gnugo are larger than the variability when compared to humans, and also when compared to each other. We have included the result obtained for games played by AlphaGo, based on the small 50-game database available \cite{alphabase}; despite the smallness of the database, the points obtained seem to be also statistically well separated from humans.
%\begin{figure}
%\begin{center}
%\includegraphics*[width=0.99\linewidth]{fidprsprsbivsec-9x9-paper-epl.eps}
%\end{center}
%\caption{Fidelity (\ref{fidelity}) (top), Ordered Vector Similarity (\ref{OVS}) (middle) and Non-ordered Vector Similarity (\ref{NVS}) (bottom) of pairs of PageRanks% as a function of the dispersion $\sigma$ (\ref{sigmadef}) of their correlator, for deterministic vs deterministic (red diamonds), Monte-Carlo vs Monte-Carlo (blue circles), Monte-Carlo vs deterministic (green triangles). For each symbol, 4 symbols correspond to 1000-game networks, one to 10000-game networks. \label{figfidsig2}} 
%\end{figure}

%%%%%%%%%%%%%%%%%%%%%%%%%%%%%%%%%%%%%%%%%%%%%%%%%%%%%%%%%%%%%%%%%%%%%%%%%%%%%%%%%%%
\section{Conclusion}
%%%%%%%%%%%%%%%%%%%%%%%%%%%%%%%%%%%%%%%%%%%%%%%%%%%%%%%%%%%%%%%%%%%%%%%%%%%%%%%%%%%

Our results show that the networks built from computer-played games and human-played games have statistically significant differences in several respects, in the spectra of the Google matrix, the PageRank vector or the first eigenvectors of the matrix. There are also differences between the different types of algorithms which can be detected statistically, from deterministic to Monte Carlo and even (although the database is smaller) the recent AlphaGo. In general, the computer has a tendency to play using a more varied set of most played moves, but with more correlations between different games for the deterministic program (Gnugo) and less for the stochastic one (Fuego). 

These statistical differences could be used to devise a Turing test for the go simulators, enabling to differentiate between the human and the computer player. Interestingly enough, it does not seem to require very large databases to reach statistical significance. We note that a manifestation of these differences was noted during the games played by AlphaGo against world champions in 2016 and 2017: the computer program used very surprising strategies that were difficult to understand by human analysts following the games.

The results shown in this study show that the computer programs simulating complex human activities proceed in a different way than human beings, with characteristics which can be detected with statistical significance using the tools of network theory. It would be very interesting to probe other complex human activities with these tools, to specify if the differences between human and computers can be quantified statistically, and to deduce from it the fundamental differences between human information processing and computer programming.

\acknowledgments We thank Vivek Kandiah for help with the computer programming and scientific discussions. We thank Calcul en Midi-Pyr\'en\'ees (CalMiP)  for  access  to  its  supercomputers. OG thanks the LPT Toulouse for hospitality.

\end{document}